\documentclass[useAMS,usenatbib]{mn2e}
\usepackage{graphicx}

\newcommand{\msun}{${\rm M_{\sun}}$}

\def\ltsima{$\; \buildrel < \over \sim \;$}
\def\simlt{\lower.5ex\hbox{\ltsima}}
\def\gtsima{$\; \buildrel > \over \sim \;$}
\def\simgt{\lower.5ex\hbox{\gtsima}}
\def\kpc{{\rm\,kpc}}

\def\msun{{\rm\,M_\odot}}

\def\pc{{\rm\,pc}}

\newcommand{\apj}{ Astrophys. J.}
\newcommand{\apjl}{Astrophys. J. Let.}
\newcommand{\apjs}{Astrophys. J. Suppl.}
\newcommand{\aj}{Astron. J.}
\newcommand{\mnras}{Mon. Not. R. Astron. Soc.}
\newcommand{\asa}{A\&A}

\newcommand{\pasp}{Pub. Astron. Soc. Pac.}

\makeatletter
\newcommand\ion[2]{#1$\;${\small\rmfamily\@Roman{#2}}\relax}%
\makeatother

\def\s{\ifmmode \widetilde \else \~\fi}
\def\={\overline}

\def\spose#1{\hbox to 0pt{#1\hss}}

\def\lta{\mathrel{\spose{\lower 3pt\hbox{$\mathchar"218$}}
     \raise 2.0pt\hbox{$\mathchar"13C$}}}
\def\gta{\mathrel{\spose{\lower 3pt\hbox{$\mathchar"218$}}
     \raise 2.0pt\hbox{$\mathchar"13E$}}}
\def\Dt{\spose{\raise 1.5ex\hbox{\hskip3pt$\mathchar"201$}}}    
\def\dt{\spose{\raise 1.0ex\hbox{\hskip2pt$\mathchar"201$}}}    

\def\dotsfill{\leaders\hbox to 1em{\hss.\hss}\hfill}

\loadboldmathitalic 
\title[Substructure in NGC~5128]
{Are small-scale sub-structures a universal property of galaxy halos?
The case of the giant elliptical NGC~5128}
\author[M. Mouhcine, R. Ibata \& M. Rejkuba]
{M. Mouhcine$^{1}$, R. Ibata$^{2}$, M. Rejkuba$^{3}$\\
$^1$ Astrophysics Research Institute, Liverpool John 
      Moores University, Twelve Quays House, Egerton 
      Wharf, Birkenhead, CH41 1LD, UK\\
$^2$ Observatoire Astronomique de Strasbourg (UMR 7550),
      11, rue de l'Universit\'e, 67000 Strasbourg, France\\
$^3$ ESO, Karl-Schwarzschild-Strasse 2, 
      D-85748 Garching, Germany  }

\date{\today}
\begin{document} 
\maketitle 
\begin{abstract}
We present an analysis of the spatial and chemical sub-structures in a remote halo field 
in the nearby giant elliptical galaxy Centaurus A (NGC~5128), situated $\sim 38\kpc$ 
from the centre of the galaxy. The observations were taken with the Advanced Camera 
for Surveys instrument on board the Hubble Space Telescope, and reach down to the 
horizontal branch. In this relatively small $3.8\kpc \times 3.8\kpc$ field, after correcting 
for Poisson noise, we do not find any statistically strong evidence for the presence of 
small-scale sub-structures in the stellar spatial distribution on scales $\ga 100$\,pc. 
However, we do detect the presence of significant small spatial-scale inhomogeneities 
in the stellar median metallicity over the surveyed field. We argue that these localized 
chemical substructures could be associated with not-fully mixed debris from the 
disruption of low mass systems. NGC~5128 joins the ranks of the late-type spiral 
galaxies the Milky Way, for which the stellar halo appears to be dominated by small-scale 
spatial sub-structures, and NGC~891, where localized metallicity variations have been 
detected in the inner extra-planar regions. This suggests that the presence of small-scale 
sub-structures may be a generic property of stellar halos of large galaxies.
\end{abstract}


\section{Introduction}

\footnotetext[1]{This work was based on observations with the NASA/ESA Hubble Space
Telescope, obtained at the Space Telescope Science Institute, which is operated by the 
Association of Universities for Research in Astronomy, Inc.,under NASA contract 
NAS 5-26555.}

The early history of galaxy assembly leaves its signatures in the properties of the 
stellar content of the outer regions of galaxies. The currently-favored scenario of 
galaxy formation in a hierarchical context predicts that the vast majority of stars in 
the outskirts of galaxies should be accreted from disrupted satellite galaxies 
\cite[e.g.][]{abadi06,zolotov09}. The ages, metallicities, spatial distribution, and 
amount of sub-structures in the faint outskirts of present-day galaxies are therefore 
directly related to their assembly histories, and to issues such as the suppression 
of star formation in small halos. 

Recent wide field observations of the Milky Way have led to the identification of a 
number of large-scale stellar features, attributed either to past accretion of identified 
satellites \citep[e.g.][]{ibata94,yanny00,yanny03,newberg02,majewski03}, others of 
unknown progenitors \cite[e.g.][]{grillmair06,belokurov07}, and others to the disruption 
of globular clusters \citep[e.g.][]{odenkirchen01,grillmair09}. The accretion of stars 
from satellites is clearly a contributor to the stellar haloes of galaxies. Quantitative analysis 
of wide-field imaging data from the Sloan Digital Sky Survey indicates that the stellar 
halo of the Milky Way is highly structured on small scales \citep{Bell08}.

For the nearest Milky Way analogue NGC~891, the analysis of deep and high resolution 
imaging data sampling a quadrant running parallel to the high surface brightness disk 
of the galaxy, and extending outward to more than 10 kpc from the plane of the galaxy, 
identified strong evidence for the presence of highly significant small-scale substructures 
in the median metallicity map \citep{ibata09}. The degree of substructures present on 
the stellar density distribution in the halo of NGC~891 is found to be strikingly comparable 
to what is present in the Milky Way \citep{ibata09}. The extra-planar regions of NGC~891 
appear to be composed of a large number of incompletely-mixed sub-populations.
The discovery of a giant stream that loops around the galaxy and a series of low surface 
brightness features in its outskirts demonstrates that NGC~891 had an active accretion 
history \citep{mouhcine10a}.

Being the nearest giant elliptical galaxy, NGC~5128 and its outer regions in particular 
have been targeted extensively as a unique testing ground for stellar population and 
formation models of elliptical galaxies \citep{soria96,harris99,marleau00,harris00,harris02,
rejkuba01,rejkuba02,rejkuba05}. 
The colour-magnitude diagrams of the stellar populations in the remote regions of the 
galaxy suggest that those regions are dominated by moderately metal-rich stars, 
i.e., ${\rm [M/H]\sim -0.65}$, with a very wide red giant branch, suggesting that stars at 
large radii span a wide range of metallicities \citep{harris99,harris02}. Different stellar
age indicators, i.e., the red clump, the asymptotic giant branch bump, the period 
distribution of long period variables, globular clusters, suggest that the outer regions 
of NGC~5128 are populated by old stars, with a likely presence of a small fraction of 
intermediate age stars \citep{soria96,rejkuba01,rejkuba03,woodley10}.

Possible formation histories of elliptical galaxies come generically in three flavours, 
first an early and rapid in-situ formation \citep[e.g.][]{larson74,harris95}, secondly 
later mergers of pre-existing disk galaxies \citep[e.g.][]{toomre77,ashman_zepf92}, 
and finally as the results of multiple dissipationless mergers and accretions of distinct 
protogalactic fragments \citep[e.g.][]{cote98}. Significant effort has been devoted over 
the years to test these scenarios using a variety of discriminants, e.g. stellar ages, 
chemical abundances, kinematics, and globular cluster systems. 
These different formation histories should give rise also to different stellar halo 
structures. The in-situ formation would predict relatively few substructures, as 
the formation epoch was many dynamical times ago. The stellar halos of elliptical 
galaxies that have grown hierarchically through mergers and accretions would be 
however substantially lumpy, and possibly as structured as those predicted by the 
most extreme cold-dark matter hierarchical formation models.

The purpose of the present paper is to examine the distribution of the stellar populations 
in the outer regions of NGC~5128, and to investigate in particular the clumpiness of the 
spatial and chemical structure of the halo population of a giant early-type galaxy for the 
first time. The data upon which this study is based have been presented in detail in 
\citet[][hereafter referred to as R05]{rejkuba05}. The layout of this paper is as follows: 
in \S\ref{data} we briefly introduce the data set used in the paper, while \S\ref{analysis} 
presents the analysis to uncover small-scale sub-structures over the surveyed area. 
The implications of our findings are discussed in \S\ref{discussion}, and we draw our 
conclusions in \S\ref{conclu}.
 
Throughout this paper we assume a distance of 3.8~Mpc, i.e., distance modulus of 
$(m-M)_0 = 27.92$ \citep[][]{rejkuba05,harris09}. We note that NGC~5128 is located at 
relatively low Galactic latitude ($\ell=309.52^\circ$, $b=19.42^\circ$), and therefore the 
targeted field suffers from significant (though not large) extinction from foreground dust. 
Following R05, we adopt a reddening value of $E(B-V)= 0.11$ \citep{Burstein82, schlegel98}, 
which corresponds approximately to $A_V=0.36$ and $A_I=0.21$.

\begin{figure}
\begin{center}
\includegraphics[angle=-90,width=\hsize]{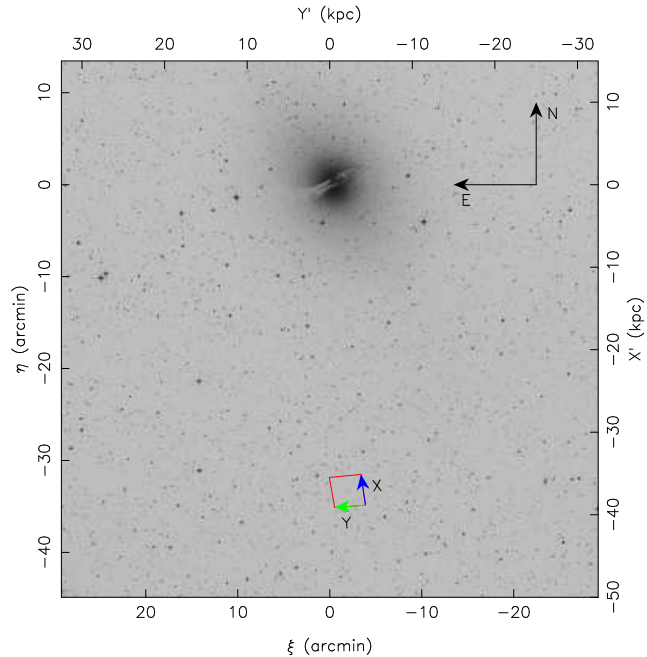}
\end{center}
\caption{The location of the ACS field studied in this contribution lies $\sim 38$\,kpc to 
the South of NGC~5128, superimposed on a photographic image extracted  from the Digitized Sky 
Survey. The scale of the region is shown both in arcmin and kpc at the distance of NGC~5128. 
The $X$ and $Y$ axes of the ACS mosaic, as constructed by \citet{rejkuba05} are displayed 
on the image; these correspond approximately to the directions North and East, respectively.}
\label{pointing}
\end{figure}

\begin{figure*}
\begin{center}
\hbox{
\hskip -0.25pc
\includegraphics[angle=0,width=20pc]{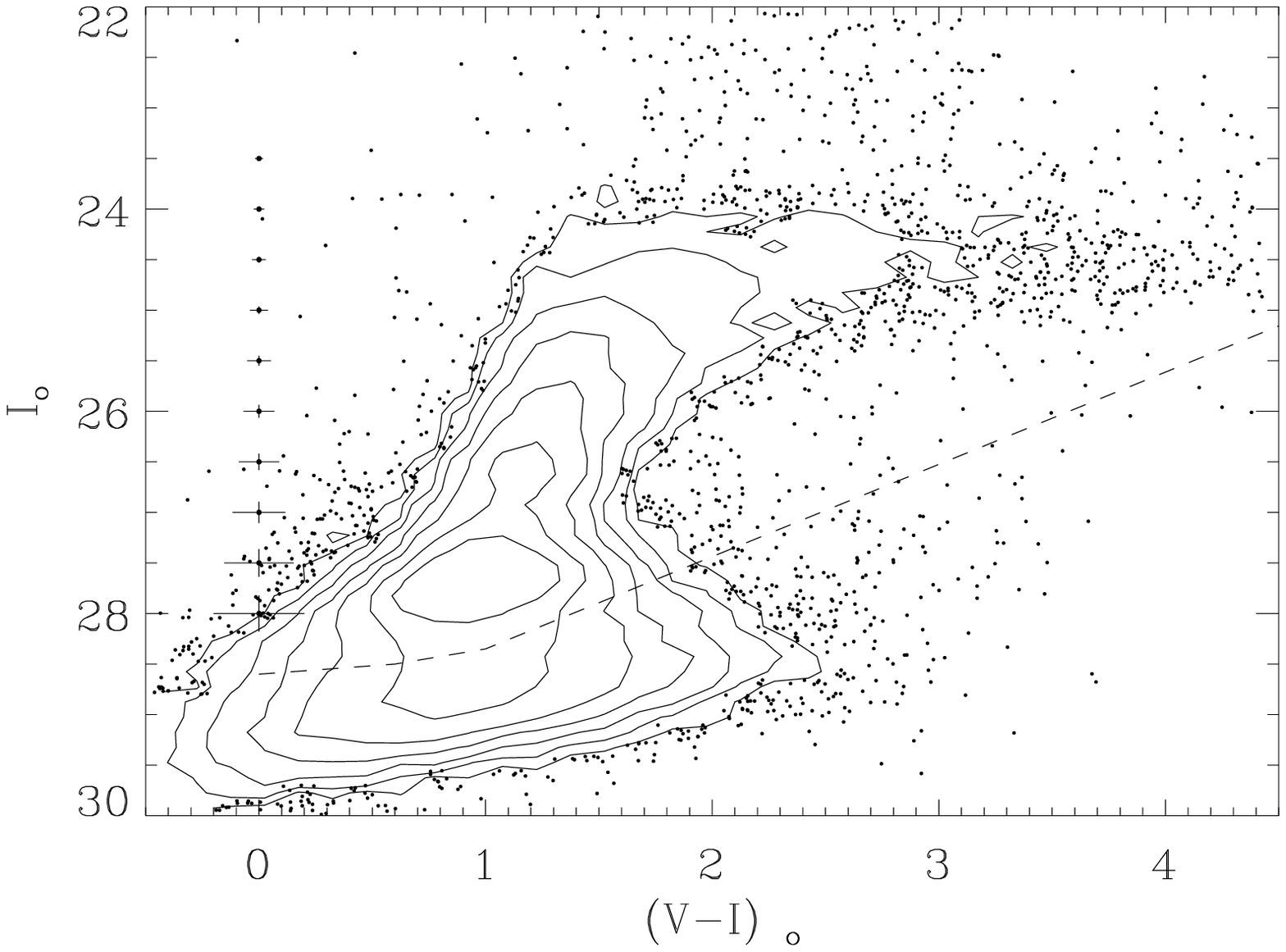}
\hskip 1.75pc
\includegraphics[angle=0,width=20pc]{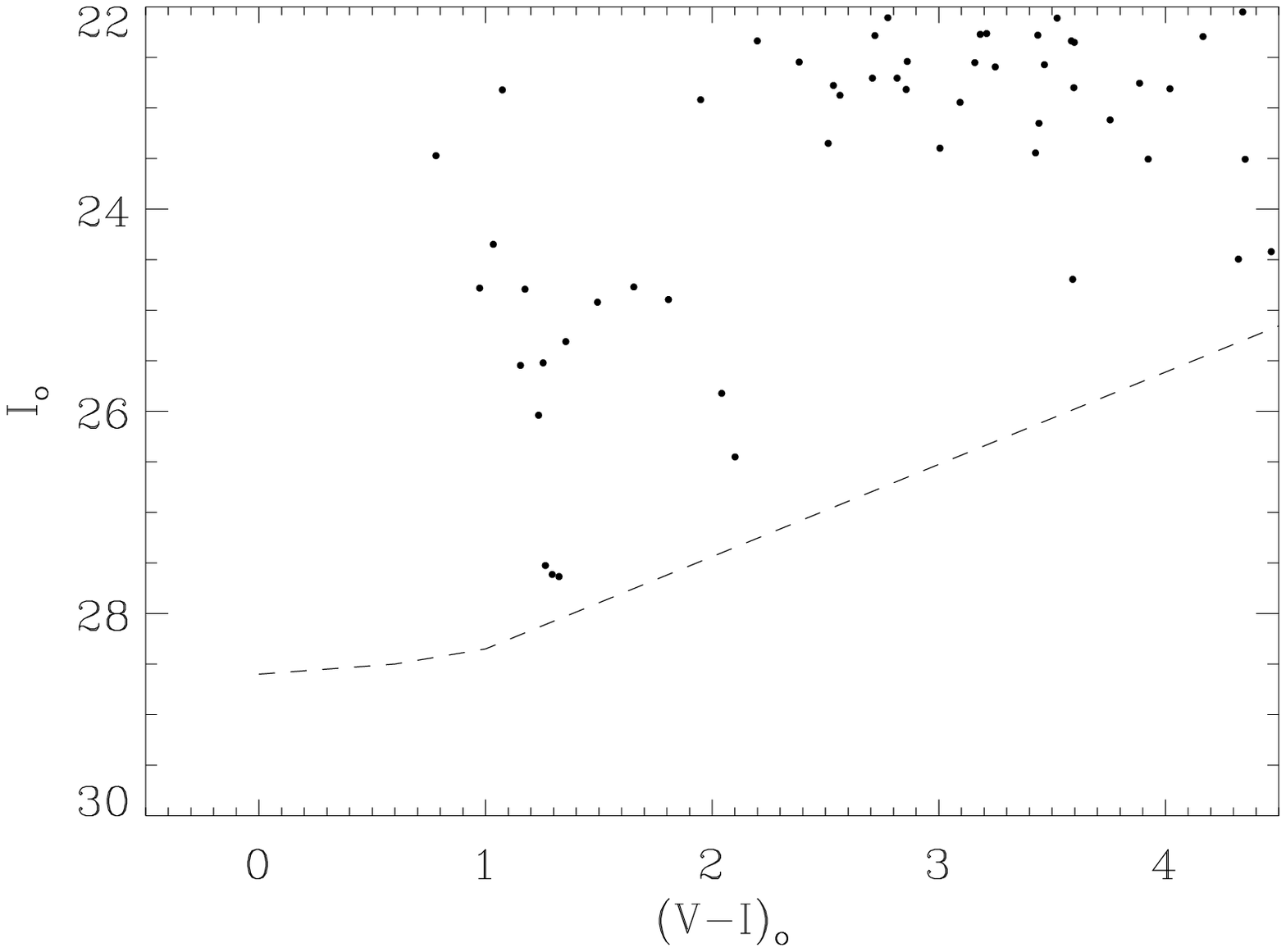}
}
\end{center}
\caption{The left panel shows the foreground reddening-corrected colour-magnitude diagram 
of the targeted halo field. The dashed line shows the 50\% detection completeness level. 
The right panel shows the colour-magnitude diagram of foreground stars as predicted by the
Besan\c{c}on Galactic population model in the direction of NGC~5128 over the same field-of-view
as covered with a single ACS pointing. }
\label{cmds}
\end{figure*}

\begin{figure*}
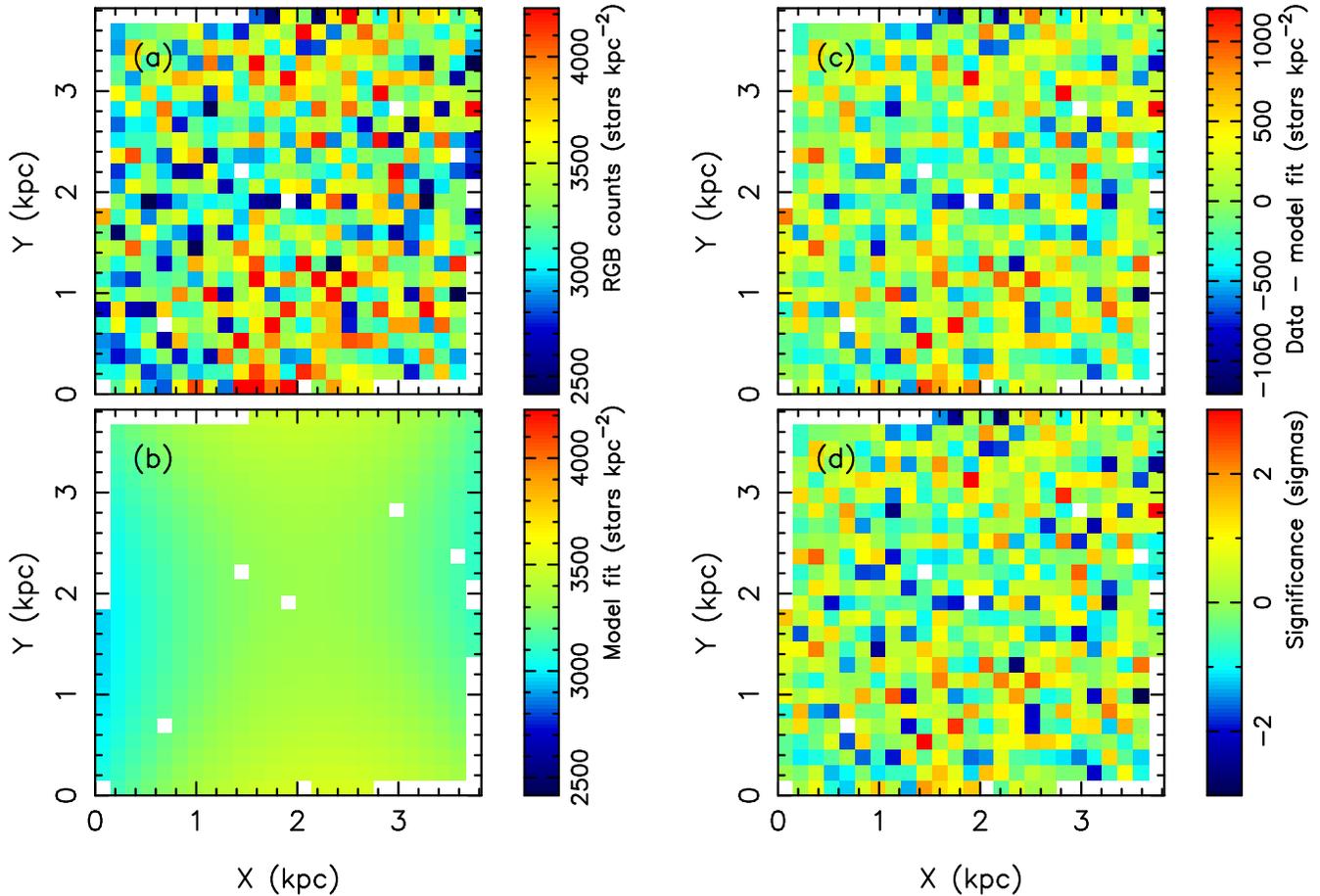

\begin{center}
\hbox{
\hskip -0.25pc
\includegraphics[angle=0,width=20pc]{N5128_halo.fig03a.ps}
\hskip 1.75pc
\includegraphics[angle=0,width=20pc]{N5128_halo.fig03c.ps}
}
\end{center}
\caption{Panel `a' shows the surface density of RGB stars in the ACS field in the coordinate 
system of Fig.~1, binned into a $25\times25$ super-pixel array. Clearly, at this distance 
from the centre of the galaxy, the density is approximately uniform. Fitting a smooth 
two-dimensional model to the data (panel `b' --- fitting details are given in the text) and
subtracting shows the small-scale variations apparent in panel `c'; the significance of these 
variations is displayed on panel `d'.}
\label{sub_density}
\end{figure*}

\section{Data}
\label{data}

As described in detail in R05, a single deep pointing was observed with the Advanced Camera 
for Surveys (ACS) camera on-board the Hubble Space Telescope (HST) in the halo of the galaxy 
NGC~5128 (program GO-9373), at a location approximately $\sim 38\kpc$ South of the nucleus 
(Fig.~\ref{pointing}). The target field was chosen to avoid any peculiarities of the galaxy, such as 
jet-induced star-forming regions \citep{mould00,rejkuba02}, shells \citep{malin83}, and dust lanes 
\citep{stickel04}. The F606W and F814W passbands were used (approximately broad-band 
Johnson $V$ and Kron-Cousins $I$), with a total exposure of 8.58~hr (12 full-orbits) in each filter, 
which reaches F606W$\sim 30$, F814W$\sim 29$. ALLFRAME software \citep{stetson94} has 
been used to measure the photometry of all detected sources on the ACS images. 
To select stellar objects, we applied the criteria of R05 (summarised in their Fig.~3), namely 
those sources with ALLFRAME sharpness parameter within their defined hyperbolic envelope, 
and with chi-squared $< 3$. The left panel of figure \ref{cmds} shows the foreground 
reddening-corrected colour-magnitude diagram (CMD) of all objects detected in our images and 
classified as stars in the deep ACS halo field. The CMD are shown as density contours to reveal 
features in otherwise crowded regions. In addition to the wide sequence of red giant stars, red 
clump stars are visible at $I_{\circ} \sim 27.5$ and $(V-I)_{\circ}\sim 1.0$, and those on the AGB 
bump at $I_{\circ} \sim 26.5$ and $(V-I)_{\circ}\sim 1.3$.

Careful analysis of the completeness corrections and photometric uncertainties has been 
conducted via Monte-Carlo simulations. Artificial stars were added to the frames, which were 
then reprocessed in the same way as the original data, locating the 50\% completeness level 
of the data at $I\sim 28.8$ and $V\sim 29.7$. 
The distribution of photometric differences between input and recovered magnitudes as a function 
of magnitude shows that the photometry is excellent with virtually no bias down to very faint 
magnitudes; for instance, at $I=27$ the uncertainty is $\sigma_I \sim 0.1$. We have fitted the 
R05 uncertainty functions with the following relations: 
\begin{eqnarray}
\sigma_V(V) &=& 2.883 \times 10^{-9} \exp(0.592 V + 0.672) \, , \\
\sigma_I(I) &=& 3.029 \times 10^{-7} \exp(0.530 I - 1.548) \, ,
\end{eqnarray}
which give a good approximation to the uncertainty estimates for $V<29$ and $I < 28$. 
In the analysis below we will assume that the $I$ and $V$ uncertainties are uncorrelated 
(i.e. with no colour terms, which is a reasonable assumption), so we adopt the above fitting 
functions to assign uncertainties to the individual stars in the survey. To investigate the 
properties of the stellar spatial distribution, we select only stars with excellent photometry, 
i.e., $\sigma_{\rm F606W} < 0.2$ and $\sigma_{\rm F814W} < 0.2$. We further limited the 
selection to stars with $-0.5 < (V-I)_{\circ} < 4.0$, yielding a catalogue of approximatively 
43\,400 stars.

As well as encompassing stars in the outskirts of NGC~5128, the deep ACS observations 
also intersect the foreground Galaxy. To estimate the foreground counts, we used the 
Besan\c{c}on Galactic population model \citep{robin03} to estimate the foreground 
contamination. The model predicts 70-80 stars in the range of $I_{\circ}=21-28$ over a 
field-of-view comparable to that of ACS in the direction of NGC~5128. The right panel 
of figure \ref{cmds} shows a typical predicted CMD of the Galaxy foreground stars. 
The model predicts about 15-20 Galaxy foreground stars with colours and magnitudes 
similar to those of red giant branch stars at the distance of NGC~5128. This constitutes
much less than one per cent of the total number of stellar objects classified as red giant 
stars in our photometric catalogue, indicating that the foreground contamination is not an 
issue for the analysis presented below.

Large background galaxies, as well as bright foreground stars cause holes in the stellar 
distribution map. To correct for this, and for the gap between the two CCDs in the camera, 
when studying the two-dimensional distribution of stellar density and metallicity, a mask 
was constructed by choosing suitably large elliptical areas around these problematic 
regions; in all, 3\% of the image was discarded in this manner. Only those stars outside 
of the masked regions were kept in the final catalogue.

It is well known that the photometric properties of old red giant branch (RGB) stars are
linked to their metallicities, offering therefore a way to determine the metallicity distribution 
of old stellar populations by comparing their colour-magnitude diagrams to RGB tracks.
The metallicity of stars on the red clump cannot be estimated however by comparing 
their photometry to RGB tracks; we therefore imposed a faint cut-off at $M_I < -0.75$ 
to the catalogue selected above in order to retain only bona-fide RGB stars. This limit 
corresponds to $I_{\circ} < 27.17$, or $I < 27.38$, where the $I$-band measurement 
uncertainty is $\sigma_I=0.13$. An additional quality cut of $\sigma_V=0.13$ was 
imposed to ensure that the $V$-band measurements were also of good quality, yielding 
a catalogue of about 13\,300 RGB stars. It is worth mentioning that these magnitude cuts 
do not exclude metal-rich RGB stars. To estimate the metallicities of the selected RGB 
stars, we repeat the procedure explained in \citet{mouhcine07}, interpolating between 
the models of \citet{vandenberg06} of $\alpha$-enhanced, i.e., [$\alpha$/Fe]=0.3, RGB 
stars of mass $0.8\msun$. The models span ${\rm -2.014 < [M/H] < -0.097}$ approximately 
in steps of 0.1 dex, and are complemented at the metal rich end by two models with 
${\rm [M/H] = 0.0}$ and ${\rm [M/H] = +0.4}$. Stars outside of the colour-magnitude 
range covered by these RGB tracks were flagged and not used in subsequent analysis 
(i.e., no extrapolation beyond the validity of the models was attempted). At the distance 
of NGC~5128, the asymptotic giant branch (AGB) bump is found to be located at 
$I\sim 26.8$ with a dispersion of 0.12 mag (see R05 for more details). 
The sample of selected bona-fide RGB stars therefore contains AGB bump stars. 
To examine the effects of this on the estimates of the chemical properties of the 
population of RGB stars, we constructed two samples by selecting stars brighter than 
$I=27.38$ and $I=26.6$ respectively. The metallicity distribution functions of both samples 
are found to be almost identical, i.e., the overall shape is unchanged and the changes in 
the median and/or average metallicities are much smaller than the typical estimate of the 
metallicity error. This indicates that the presence of AGB bump stars in the selected sample 
of RGB stars does not bias the results discussed below.

\section{Analysis}
\label{analysis}

\subsection{Spatial sub-structure}

We first perform an analysis of the number density variations over the area of the ACS 
survey. Panel `a' of Fig.\,\ref{sub_density} shows the star-count density map in the $X,Y$ 
CCD coordinate system chosen by R05. We opted to bin the stars in a grid of $25\times25$ 
super-pixel, to contain $\sim 100$ sources (to ensure a signal-to-noise ratio of $\sim 10$). 
The star number density has been corrected for the mask, and only super-pixels with 
a usable surface area larger than 50\% were kept. The map appears approximately
homogenous, with no obvious gradient. This is not very surprising since the observed 
field covers a small physical size at the distance of NGC~5128 compared to its projected 
radial distance. Since we do not have an a-priori model of the large-scale stellar distribution 
over the targeted field, we fit the observed stellar distribution with a two-dimensional 
Legendre polynomial function (shown in panel `b'), with up to quadratic terms in $X$ 
and $Y$ including cross-terms (i.e. 6 parameters), which is sufficient to provide a smooth 
empirical template to compare the data to. The residuals between the data and the best 
smooth model are shown in panel `c', along with their statistical significance in panel `d', 
where we have assumed only Poisson noise uncertainties (the effect of the mask is properly 
accounted for). There appears to be no coherent large-scale structure in the residuals that 
might indicate the presence of, for instance, streams or shells of stellar debris.

How likely are these small-scale variations about the model of the stellar spatial distribution? 
A way to quantify the significance of the presence of substructure in a survey is to estimate 
the rms deviation of the data around a smooth model, taking into account the expected 
Poisson noise in the model. \citet{Bell08} proposed the following statistic:
$\sigma/{\rm total} = { {\sqrt{ {{1}\over{n}} \sum_i (D_i - M_i)^2 - {{1}\over{n}} \sum_i (M_i^\prime - M_i)^2}}/ {{{1}\over{n}} \sum_i D_i}}$, 
where $n$ is the number of discrete bins the sample is divided into, $D_i$ are the individual 
observed counts in those bins, $M_i$ are the model values in the pixels, and $M_i^\prime$ 
is a Poisson realization of the model with mean $M_i$. Thus the numerator is the pixel scatter 
of the data around the model minus the expected scatter in the model, while the denominator 
was chosen to be the total of the Poisson scatter. This statistic is convenient because it is 
independent of the choice of binning scale (provided that the substructures are well sampled
by the chosen binning scale), and the number of sources in the survey, and furthermore the 
Poisson noise contribution is removed. 

For a bin size ranging from 100\,pc to 350\,pc, i.e., grids of $42\times 42$ to $12\times 12$
respectively, the fractional rms deviation $\sigma/{\rm total}$ varies from 0.04 to 0.02 with 
probabilities of finding those values of this statistic by chance ranging from 12\% to 22\%, 
assuming the smooth model shown in panel `b' of Fig.\,\ref{sub_density}. 
This clearly indicates that the small-scale deviations of the data around the model shown 
in Fig.\,\ref{sub_density} are consistent with what would be expected from Poisson-noise 
deviations from a smooth model. The stellar populations within the surveyed field are 
then most likely smoothly distributed at physical scales larger than $\sim 100$\,pc. 

This contrasts with what was reported for the two large late-type galaxies where the 
properties of small-scale substructures have been investigated so far. 
\citet{Bell08} found that the fractional rms deviation exceeds $\sigma/{\rm total} = 0.4$ 
in the Galactic halo region probed by the SDSS. For the Milky Way analogue NGC~891, 
\citet{ibata09} have estimated $\sigma/{\rm total} = 0.14$ for the extra-planar regions 
within $\sim 10$\,kpc from the galactic plane, with an extremely small probability of 
finding this value by chance. This clearly indicates that the spatial distribution of halo 
stars in both large late-type spiral galaxies is most likely dominated by significantly 
lumpy components. 

The catalogue used in the present analysis integrates stellar populations along the line 
of sight through the entire halo of the galaxy. The contrast introduced by any sub-structure 
will be then reduced, especially in the presence of multiple sub-structures along the 
probed line of sight. It should then be appreciated that the low fractional rms deviation 
found here does not necessarily imply that localized spatial over-densities are absent 
in the halo of NGC~5128. Furthermore, the present survey samples a halo region much 
smaller than that studied in NGC~891 by \citet{ibata09}, and a tiny fraction of the halo 
when compared to the vast region of the Galactic halo sampled by the SDSS data 
examined in \citet{Bell08}. Wide-field observations covering a representative volume of 
the outskirts of the galaxy are needed to establish (or to rule out) firmly the predominance 
of a smooth stellar component at large radii in NGC~5128, and to quantify comprehensively 
the amount of sub-structures in those regions.

\subsection{Metallicity sub-structures}

In this section we investigate the properties of the two-dimensional distribution 
of stellar metallicities over the surveyed region. Given the smaller size of the sample
of stars for which we can measure photometric metallicities, we decided to distribute 
stars in a $12\times12$ bin grid ($350\pc \times 350\pc$) to ensure a high 
signal-to-noise ratio in each super-pixel. The star number density has been corrected 
for the mask, and we have discarded the few super-pixels where the usable surface 
area was less than 50\%. 
As in \citet{ibata09}, we use the median metallicity as a statistic to investigate the 
spatial variations of stellar populations in the outer regions of the galaxy. This choice 
presents the advantages of being insensitive to e.g. variations of the population spatial 
density, holes due to bright stars, etc. 

\begin{figure}
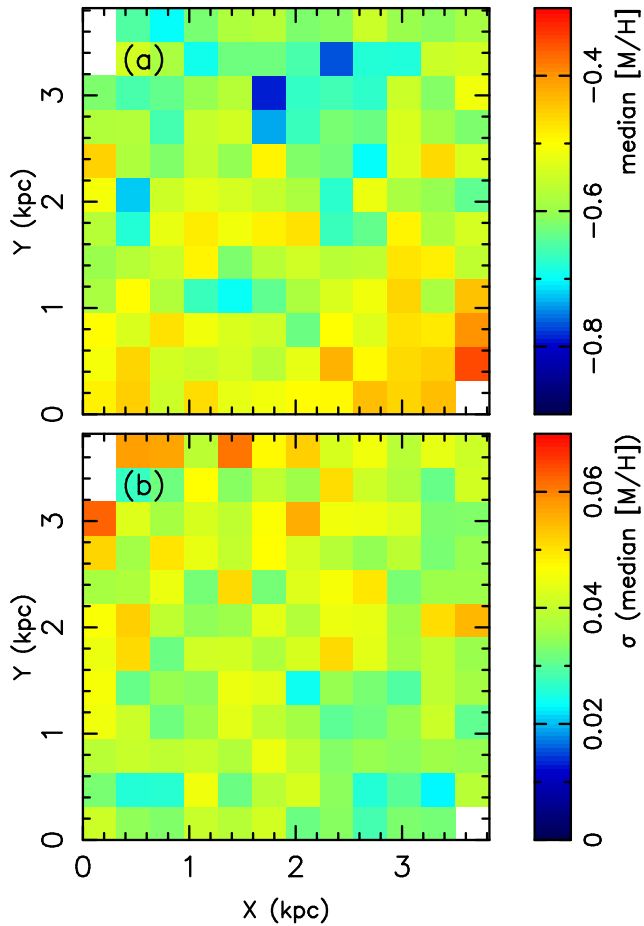

\begin{center}
\includegraphics[angle=0,width=\hsize]{N5128_halo.fig04a.ps}
\vskip 0.125pc
\includegraphics[angle=0,width=\hsize]{N5128_halo.fig04b.ps}
\end{center}
\caption{Panel `a' displays the median metallicity calculated in each of the spatial 
super-pixels. A slight gradient is apparent, with the top left-hand corner being more 
metal-poor than the bottom right. The corresponding uncertainties, calculated from 
Monte-Carlo simulations accounting for both the photon noise and the sampling 
errors, are shown on the panel `b'.}
\label{medianZ}
\end{figure}

The map of median metallicities estimated in the super-pixels is shown in the 
upper panel of Fig.\,\ref{medianZ}. The lower panel of the figure shows the map of 
the uncertainties of median metallicities. The errors on the median metallicity in each 
super-pixel is estimated by simulating the effects of both the photometric uncertainties 
on the metallicity estimate for each star belonging to the super-pixel and the sampling 
uncertainty due to the limited number of stars per super-pixel 
\citep[see][for a detailed description]{ibata09}. The map shows that the combined 
random errors on the median metallicities in the super-pixels are of the order of 
$0.02-0.04$~dex.

The first clear feature of the median metallicity map is the presence of a large scale 
gradient over the field, with the average median metallicity decreasing when moving 
diagonally with increasing $Y$ (i.e., moving from the West to the East).
A second striking feature of the median metallicity map is that the pixel-to-pixel 
variations are significantly larger than the random errors. Note that while we notice 
the presence of numerous super-pixels dominated by stellar populations with lower 
median metallicities, i.e., ${\rm [M/H] \la -0.7}$, than that of the overall stellar population, 
i.e., ${\rm [M/H] \sim -0.6}$, it appears however that metal-rich, i.e., ${\rm [M/H] \ga -0.5}$, 
super-pixels are rare. How likely are those small-scale variations about a large-scale 
smooth model of the median metallicity distribution?

\begin{figure}
\begin{center}
\includegraphics[angle=0,width=\hsize]{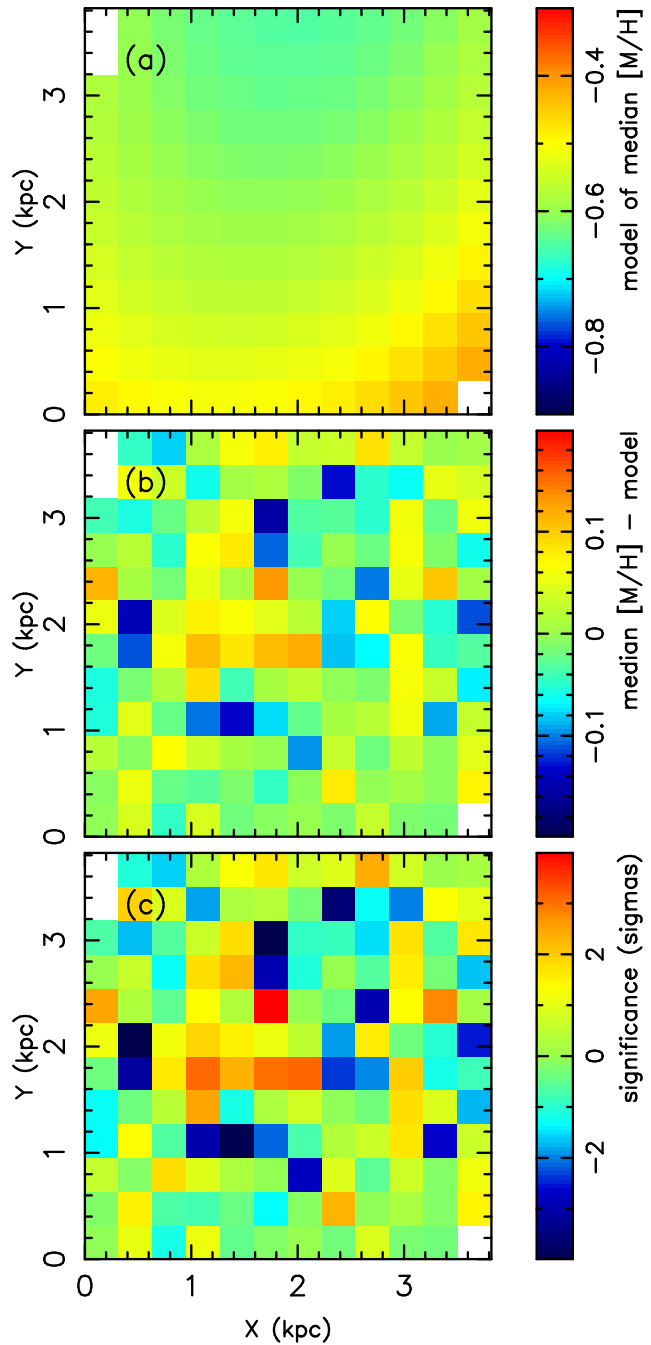}
\end{center}
\caption{A two-dimensional polynomial model fit to the median metallicity map 
(panel `a' of Fig.\,\ref{medianZ}) is shown in panel `a'. The residuals between the 
data and this model are displayed in panel `b'. Using the uncertainty estimates on 
the metallicity metallicity from panel `b' of Fig.\,\ref{medianZ}, the statistical 
significance levels of the metallicity variations are shown in panel `c'.}
\label{modelZ}
\end{figure}

To represent the large-scale spatial distribution of median metallicities, we fitted the 
observed distribution with a two-dimensional Legendre polynomial (up to order 3 in 
both spatial directions, i.e., 10 parameters). The fit is shown in the top panel of 
Fig.\,\ref{modelZ}, and exhibits clearly a large-scale median metallicity gradient. 
The middle panel of Fig.\,\ref{modelZ} shows the difference between the map of median 
metallicities and the best smooth model, while the lower panel shows the map of the 
significance levels of these differences. The lower panel of the figure shows clearly the 
presence of localized variations at the 3-5\,$\sigma$ level in the median metallicity map, 
occurring on the scale of a super-pixel or a few super-pixels, and distributed over the 
field. When comparing the significance map of the deviations of the median metallicity 
map about the best smooth model to the maps of the spatial density distribution and 
the deviation about the best smooth model (see Fig.\,\ref{sub_density}), we notice 
that the small-scale sub-structures in the median metallicity map are not systematically 
associated to small-scale stellar density sub-structures. As already discussed in detail 
in our study of NGC~891, the number density of stars required to cause statistically 
significant deviations about a smooth median metallicity map is rather small, causing 
only a modest enhancement in the stellar distribution that is much harder to detect 
from its density contrast. The stellar number density needed to cause a deviant 
super-pixel in the median metallicity map depends on the shape of the metallicity 
distribution functions of both the smoothly-distributed population and the 
super-imposed sub-structures, but in broad terms an additional stellar population
numbering $10-15$ per cent of the smooth population could induce a $\sim 0.1$ dex 
metallicity variation \citep{ibata09}. While this mass fraction of the super-imposed 
population is of the order of the change in density over a typical super-pixel, the 
induced change in the median metallicity of the super-pixels is many times larger 
than the typical errors on the estimate of the median metallicity (see the lower panel 
of Fig.\,\ref{medianZ}). It is then not surprising that sub-structures detected in the 
median metallicity map are not inducing significant deviations in the number density 
map. 

\begin{figure}
\begin{center}
\includegraphics[angle=-90,width=\hsize]{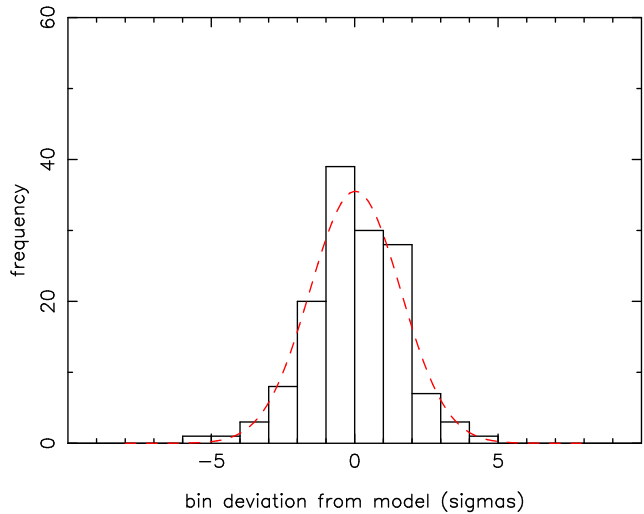}
\end{center}
\caption{The distribution of statistical significance of the deviations from the best 
smooth model of the median metallicity map shown in the top panel of Fig.\,\ref{modelZ}. 
The Gaussian model fit superimposed on this distribution has a dispersion of $1.58$. 
The distribution of significance levels has an extremely small probability of being 
drawn by chance from a Gaussian distribution of unit dispersion.}
\label{dist_sigma}
\end{figure}

The distribution function of the statistical significance of the variations about the 
best smooth model of the median metallicities map is shown in Fig.\,\ref{dist_sigma}. 
The dashed line shows a fitted Gaussian of dispersion equal to 1.58. The probability of 
finding such a large dispersion by chance (we expect unit dispersion if the uncertainties 
have been correctly estimated), given the large number of bins, is smaller than 0.1\%. 
This strongly indicates that the variations about the best smooth model of the spatial 
distribution of median metallicities are significantly larger than expected for a 
population where the median metallicity varies spatially in a smooth manner. 

Perhaps the simplest explanation of these large deviations is that we have 
underestimated the uncertainties in the median metallicity by $30-50\%$. This would 
require that the photometric errors reported by R05 to be underestimated by a similar 
factor. In our study of NGC 891 we were able to verify our uncertainty estimates by 
comparing measurements in overlapping fields. Unfortunately, we do not have the 
necessary complementary data to perform a similar check here. Although we cannot 
dismiss this possibility, experience from our NGC 891 study (and other projects) 
suggests that artificial star simulations give reliable uncertainty estimates.

The large pixel-to-pixel variations reported in the median metallicity map are unlikely 
to result from errors in the zero-point over the ACS field. To induce the observed 
small-scale metallicity changes, large variations in the zero point (of the order of 
0.1 mag or more) are required, which is extremely unlikely for the very stable 
instrument that the ACS is. 

Could the large-scale metallicity gradient and the small-scale chemical substructures 
we have uncovered be due to systematic biases? To convert the RGB photometric 
properties into metallicities, we have had to assume that the stellar populations 
dominating the surveyed field are uniformly old. Due to the so-called age-metallicity 
degeneracy, a red RGB star can be either more chemically evolved than a bluer 
one or older. In the interpolation procedure, we have assumed a single old age for 
all the stars, which may introduce a bias of up to 0.1 dex, if the age is $\sim 4$ Gyr 
younger \citep[e.g.][]{rejkuba05}. 

\begin{figure}
\begin{center}
\includegraphics[width=\hsize]{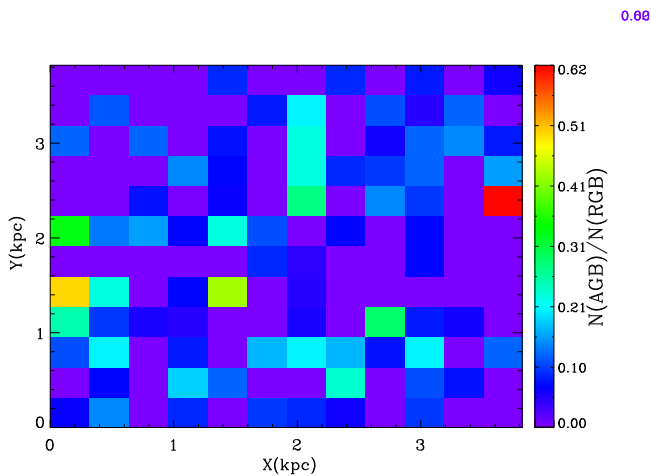}
\end{center}
\caption{The  spatial distribution of the ratio of AGB stars  to RGB stars over the ACS field.}
\label{agb_rgb_map}
\end{figure}

The AGB stars are known to be efficient tracers of intermediate age (1-6 Gyr) stellar 
populations \citep{ml02,marigo07}. The abundance of AGB stars normalized to 
RGB stars, which are tracing predominantly older populations, is a good indicator 
of a stellar population age \citep[e.g.][]{frogel90}. The map of the ratio of AGB stars, 
selected as stars with $22.0<I_{\circ}<23.5$ and $1.5<(V-I)_{\circ}<3.0$, and all other 
quality criteria identical to the RGB sample summarized in \S\ref{data}, to stars in the 
brightest one magnitude of the RGB is shown in Fig.\,\ref{agb_rgb_map}. The spatial 
distribution of the AGB to RGB ratio is fairly smooth, with no apparent spatial gradient. 
This suggests the absence of an AGB gradient and then most likely the absence of an age 
gradient across the observed field. Furthermore, no enhancements of AGB stars relative 
to RGB stars are associated to any of the super-pixels that are deviants from a smooth 
model of the median metallicity map by more than $3\,\sigma$. Interestingly, the typical 
AGB to RGB ratio, i.e., $\la 0.2$.  is consistent with an old stellar population 
\citep{mould05}. The pixel-to-pixel variations of the AGB-to-RGB ratio are significantly 
smaller than what is required for the pixel-to-pixel variations of the median 
metallicities to be due only to age differences: to account for the $\ga 0.15\,$dex 
variations in median metallicity about the best smooth model for the most deviant 
super-pixels (i.e., $\pm 3\sigma$ or more) as a result of age differences, the stellar 
populations of those super-pixels have to be dominated by $\la 6\,$Gyr stars, which is 
not supported by the morphology of the colour-magnitude diagram 
\citep[see][for a detailed discussion]{rejkuba05}.

The localized median metallicity variations are also unlikely to result from the 
effects of small-scale structures in the foreground Galactic inter-stellar medium. 
Following \citet{ibata09}, by using the reasonable assumption that reddening 
variations scale approximately linearly with mean reddening, we expect less than 
a couple hundredths of a magnitude scatter at most in colour excess from foreground 
dust. For the small-scale metallicity variations to be due solely to variation in 
foreground extinction, the required offset in colour excess are either much larger 
than the expected scatter, or, more problematically, negative in the case for deviant 
metal-poor super-pixels. Small angular scale variations in foreground dust are 
therefore unlikely to contribute significantly to the observed small-scale variations 
in median metallicity. 

\begin{figure*}
\includegraphics[angle=0,width=20pc]{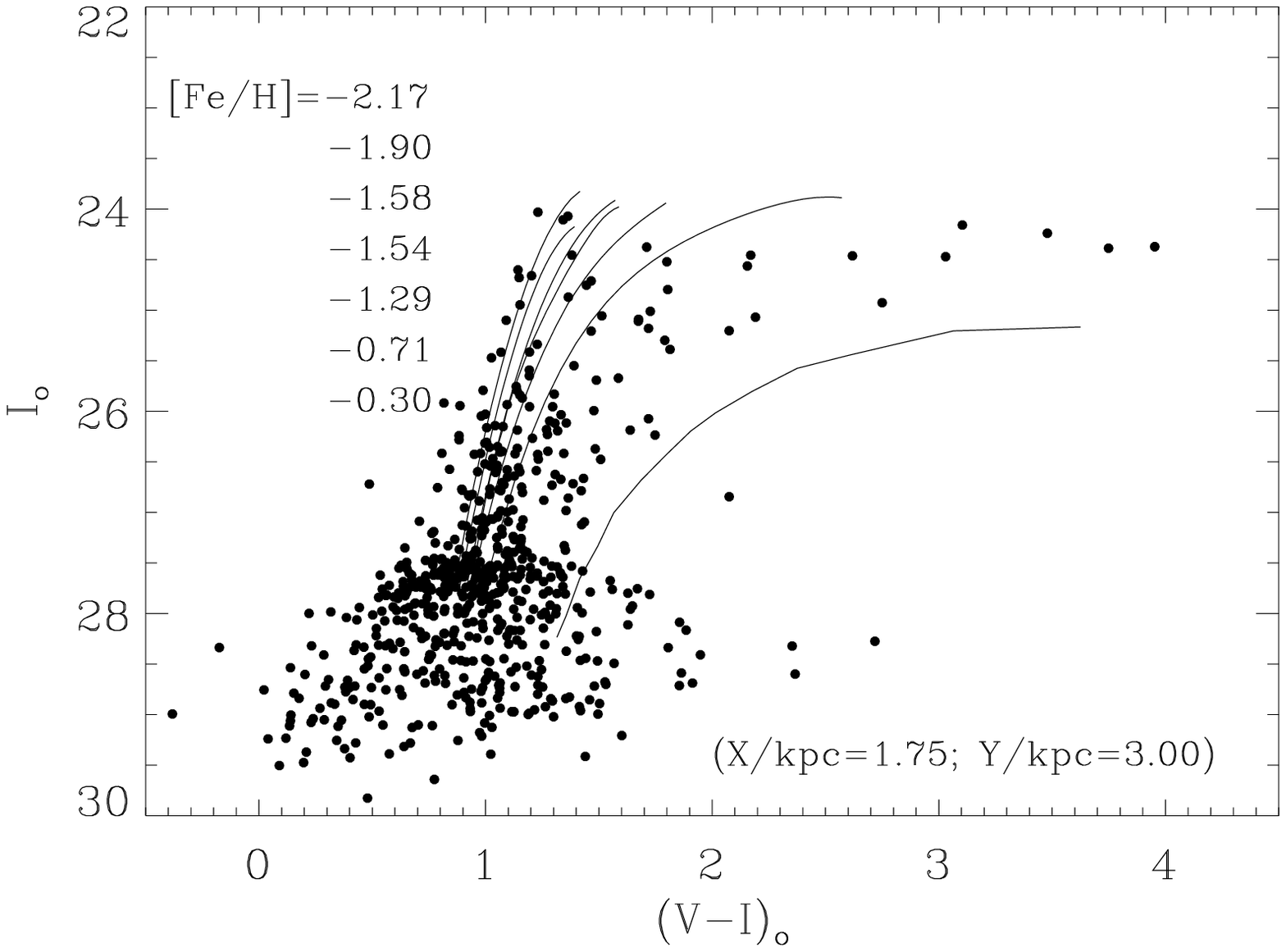}
\includegraphics[angle=0,width=20pc]{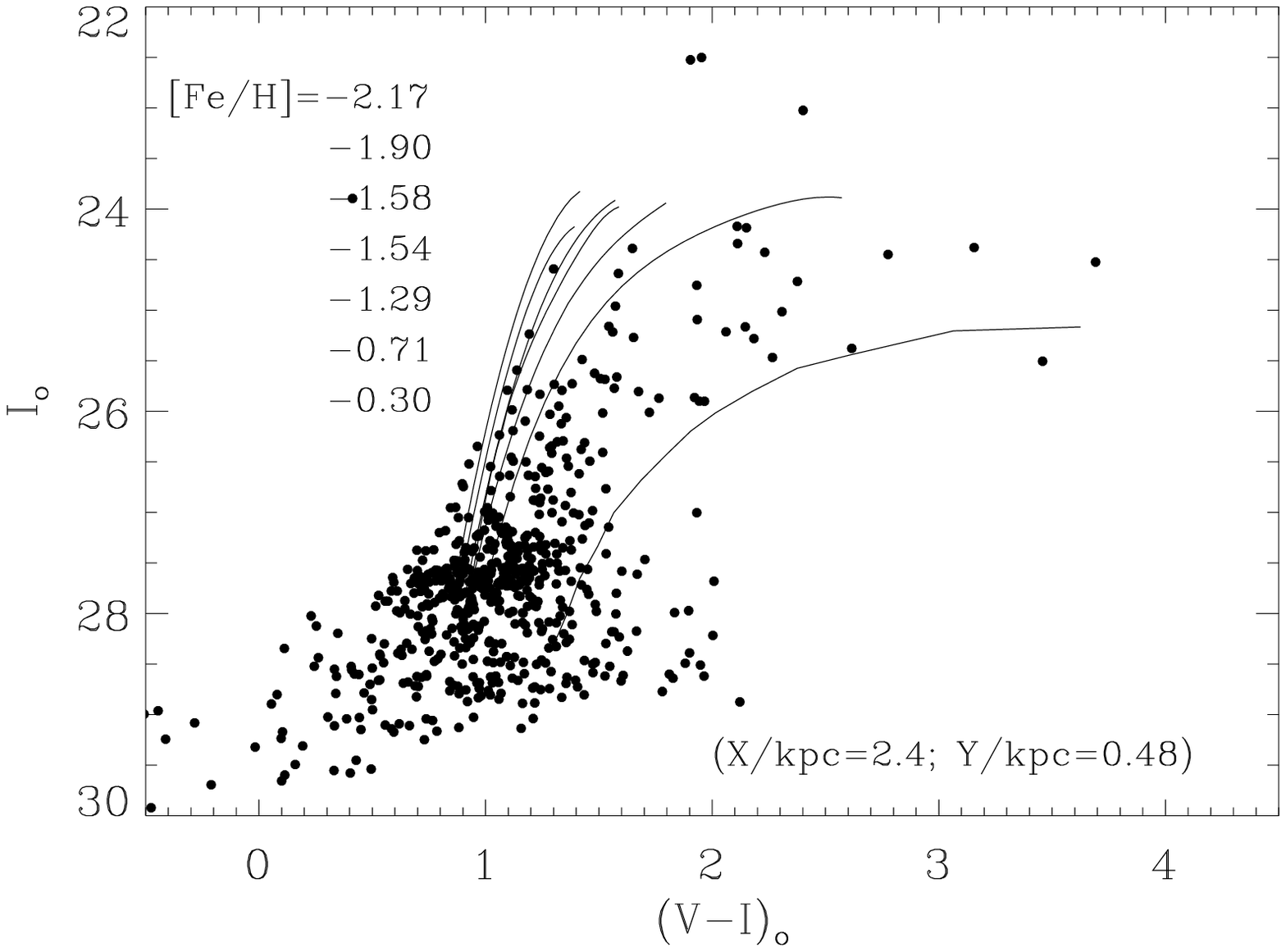}
\caption{CMDs of stars populating two super-pixels with their respective positions indicated 
in each panel. The left panel shows the CMD of the most metal-poor super-pixel in the map
shown in Fig.\,\ref{medianZ} with a median metallicity of [M/H]=-0.77, while the right panel 
show the CMD of a typical metal-rich super-pixel with a median metallicity of [M/H]=-0.41.
Note that the median metallicity for the whole field is [M/H]=-0.53. Shown also as solid lines 
are the red giant sequences of the indicated Galactic globular clusters from \citet{dacosta90} 
assuming an intrinsic distance modulus for NGC~5128 of 27.9. It is clear that the metal-poor 
side of the red giant branch is much more populated in the case of the metal-poor spatial
super-pixel than for the metal-rich one. }
\label{cmds_super_pixels}
\end{figure*}

We conclude that there is no evidence for systematic biases affecting the measure 
of stellar metallicities. So, under the assumption that the photometric uncertainties 
have been well estimated, we conclude that the statistically significant small-scale 
variations in the median metallicities of the stellar populations are present in the 
targeted halo field. To highlight the differences between thte detected small-scale 
chemical inhomogeneities, Fig.\,\ref{cmds_super_pixels} shows the CMDs of all stellar 
objects populating a metal-poor super-pixel, i.e., with a median metallicity of 
[M/H]=-0.77, and a metal-rich super-pixel, i.e., with a median metallicity of 
[M/H]=-0.47, located at (X/kpc=1.75, Y/kpc=3.00) and (X/kpc=2.40, Y/kpc=0.48) 
respectively. Fiducials of RGB sequences for Galactic globular clusters spanning a 
wide range of metallicities and shifted to the distance of NGC~5128 are overplotted. 
The stellar populations occupying both super-pixels appear to be different. 
A significant fraction of bright RGB stars populating the metal-poor super-pixel 
exhibit photometric properties consistent with metallicities in the range of [Fe/H]=-2.2 
and -1.3. Such metal-poor stars are absent in the metal-rich super-pixel. Furthermore, 
the reddest stars in the metal-rich super-pixel extend clearly to higher metallicities 
than those occupying the metal-poor super-pixel. The stellar population differences 
between the localized sub-structures in the median metallicity spatial distribution 
would suggest that the survey has revealed the presence of inhomogeneities that have 
not yet been fully blended into the smooth halo.

\section{Discussion}
\label{discussion}

The metallicity distribution function of the stellar populations in the outer regions 
of NGC~5128 has been determined over an extended range of radial distances, 
10-40 kpc \citep{harris99,harris02,rejkuba05}. The overall shape, mean metallicity, 
metallicity dispersion, fraction of metal-poor stars, are found to be strikingly 
similar \citep{rejkuba05} over these fields. This indicates that the metallicity 
gradient present within the field at $\sim 40\,$kpc (see Fig.\,\ref{modelZ}) could not 
be associated to a large-scale metallicity gradient, as it would lead to metallicities 
significantly higher than observed if extrapolated inward. The most likely explanation 
is that the median stellar metallicity gradient present within the ACS halo field is 
related to chemical inhomogeneities present locally.

Using the predictions of a $\Lambda$-cold dark matter semi-analytical galaxy 
formation model, and by summing the total star formation occurring over all the 
progenitors, \citet{beasley03} computed the metallicity distribution function of spheroid 
stars of early-type galaxies comparable in luminosity and environment to NGC~5128. 
Despite some differences, the predicted metallicity distribution function resembles 
qualitatively the observed distribution function of stellar metallicities in the outer 
regions of NGC~5128, i.e., mean metallicity, metal-rich end cut-off, and extent of 
the metal-poor tail \citep[see][]{rejkuba05}. This broad agreement suggests that the 
stellar populations in the outer regions of NGC~5128 could have been assembled 
in a hierarchical fashion. If this is indeed the case, one can expect to find the presence 
of chemical sub-structures caused by the remnants of the disruption of the predicted 
large populations of sub-halos orbiting around massive galaxies.

The localised chemical inhomogeneities could be in principle potential candidate 
satellites or remnants of past accretion events. Upon inspecting the stacked ACS 
image visually, none of the strongly deviant super-pixels shows any plausible stellar 
concentration, although clear RGB sequences are detected for all of them, apart 
from the super-pixel at (${\rm X\sim 1.8\,kpc, Y\sim 3\,kpc}$) which contains the 
faint, old, and metal-poor extended globular cluster reported in \citet{mouhcine10b}. 
The extended nature and the low luminosity of that globular cluster become very 
important when considered in light of the emerging consensus that extended clusters 
are formed preferentially in low mass galaxies \citep[e.g][]{elmegreen08,dacosta09}. 
In M31, all but one extended globular cluster in the outer halo, i.e., at projected 
distances larger than 30\,kpc, are found to be projected onto stellar substructures 
or are members of a cluster overdensity \citep{mackey10}, supporting the scenario 
where this class of globular clusters originate in low mass systems.

The properties (luminosity, surface brightness, size, central concentration) of the 
globular cluster found in the halo field resemble in every way one those of the faint 
and extended globular clusters populating the Milky Way outer halo
\citep{mouhcine10b}. This suggests that the extended cluster in the targeted field 
could have been formed in a similar environment as its Galactic counterparts, and 
experienced potentially a similar dynamical history. Strong evidence is accumulating 
that extended and faint globular clusters in the outer Galactic halo, i.e., the 
so-called young halo globular clusters, were associated with dwarf galaxies that 
have since disrupted \citep[e.g.][]{zinn93,mackey04,forbes10}. 
\citet{mackey10} have argued that the vast majority of globular clusters in the outer 
halo of M31 are most likely associated with tidal debris features observed in those 
regions, thus strongly supporting the scenario where the outer globular cluster system 
of M31 has been built up via the accretion of satellite host galaxies. It is then tempting 
to suggest that the extended faint cluster in the surveyed halo field in NGC 5128 
could have originated in a now disrupted dwarf galaxy. Interestingly, the median 
metallicity of the diffuse stellar population within the super-pixel containing the 
extended faint cluster is comparable to those measured for metal-poor super-pixels 
deviants at the $3-4\,\sigma$ levels in the observed halo field, i.e., ${\rm [M/H]\sim -0.75}$, 
suggesting that the stellar populations dominating the metal-poor super-pixels could 
be sharing the same origin. Put together, these properties provide circumstantial 
evidence that the chemical inhomogeneities detected in the observed field could 
be related to the disruption of low mass satellites.

The restricted spatial extent of the present survey limits however our ability to 
investigate the large-scale distribution of the chemical sub-structures, to quantify 
the degree of spatial sub-structures in the halo of the galaxy, to search for signs of 
the disruption of the hypothetical progenitor(s) of the detected chemical sub-structures. 
Extending the survey of the outer regions of NGC~5128 should be undertaken to tackle 
comprehensively these questions. When compared to the properties of sub-structures 
in the outskirts of late-type galaxies, it will be possible to constrain the assembly 
histories of galactic halos as a function of galaxy e.g. morphology, type, and mass.

\section{Summary \& Conclusions}
\label{conclu}

To make the first attempt to quantify the degree of substructure in the outer regions of 
elliptical galaxies, we have undertaken a structural and chemical analysis of the stellar 
content of a ${\rm 3.8\,kpc \times 3.8\,kpc}$ halo field situated $\sim 38\kpc$ from the 
centre of the nearest giant elliptical galaxy NGC~5128. The very deep colour-magnitude 
diagram of the stellar populations used in the analysis reaches the red clump. 

Comparing the observed stellar number density map of halo stars to smooth models, 
we find no statistically significant evidence for the presence of spatial density variations.
Deviations from the smooth models, quantified using the rms deviation of the data 
around the best fit model are consistent with what would be expected from Poisson-noise
deviations in a smooth model. The measured statistics appears to suggest the presence 
of a smooth stellar halo component at spatial scales larger than $\sim 100$\,pc in the
region probed by the present survey.

Estimating metallicities of RGB stars from their optical photometry, we investigated the 
spatial distribution of stellar metallicities. On top of a smoothly distributed large-scale 
map, we detect the presence of small-scale sub-structures in the median metallicity 
across the surveyed field. Although the absolute median metallicity variations are 
relatively modest, i.e., ranging from $\sim 0.08$ to $\sim 0.2$ dex, they are nevertheless 
statistically significant. Those small-scale chemical sub-structures do not appear to be 
associated with any statistically significant stellar spatial density variations. We argue 
that these subtle small-scale metallicity variations are related to accretions of low mass 
systems.



\begin{thebibliography}{}

\bibitem[Abadi et al. (2006)]{abadi06}
Abadi M.~G., Navarro J.~F., Steinmetz M., 2006, \mnras, 365, 747

\bibitem[Ashman \& Zepf (1992)]{ashman_zepf92}
Ashman K.~M., \& Zepf S.~E., 1992, \apj, 384, 50

\bibitem[Beasley et al. (2003)]{beasley03}
Beasley M.~A., Harris W.~E., Harris G.~L.~H., Forbes D.~A., 2003, \mnras, 340, 341

\bibitem[Bell {et~al.}(2008)]{Bell08}
Bell E.~F., et al., 2008, \apj, 680, 295

\bibitem[Belokurov {et~al.}(2007)]{belokurov07}
Belokurov, V., et al., 2007, \apj, 654, 897

\bibitem[Burstein \& Heiles(1982)]{Burstein82}
Burstein D., Heiles C., 1982, Astronomical Journal, 87, 1165

\bibitem[C\^ot\'e {et~al.}(1998)]{cote98}
C\^ot\'e P., Marzke R.~O., West M.~J., 1998, \apj, 501, 554

\bibitem[Da Costa \& Armandroff (1990)]{dacosta90}
Da Costa G.~S., Armandroff T. E., 1990, \aj, 100, 162

\bibitem[Da Costa et al. (2009)]{dacosta09}
Da Costa G.~S., Grebel E.~K., Jerjen H., Rejkuba M., Sharina M.~E., 2009, \aj, 137

\bibitem[Elmegreen (2008)]{elmegreen08}
Elmegreen B.~G., 2008, \apj, 672, 1006

\bibitem[Forbes \& Bridges(2010)]{forbes10}
Forbes D.~A., \& Bridges T., 2010, \mnras, 404, 1203

\bibitem[Frogel et al. (1990)]{frogel90}
Frogel J.~A., Mould J., Blanco V.~M., 1990, \apj, 352, 96

\bibitem[Grillmair \& Dionatos (2006)]{grillmair06}
Grillmair C. J. Dionatos, O. 2006, \apj, 643, L17

\bibitem[Grillmair (2009)]{grillmair09}
Grillmair C. J., 2006, \apj, 693, 1118

\bibitem[Harris et al (1995)]{harris95} 
Harris W.E., Pritchet C.~J., McClure R.~D., 1995, \apj, 441, 120

\bibitem[Harris et al.(1999)]{harris99}
Harris G.L.H., Harris, W.E., \& Poole, G.B., 1999, \aj, 117, 855

\bibitem[Harris \& Harris (2000)]{harris00} 
Harris G.L.H., \& Harris W.E., 2000, \aj, 120, 2423

\bibitem[Harris \& Harris (2002)]{harris02} 
Harris W.E., \&  Harris, G.L.H. 2002, \aj, 123, 3108

\bibitem[Harris et al.(2009)]{harris09}
Harris, G. L. H., et al., 2009, axXiv0911.3180

\bibitem[Ibata {et~al.}(2005)]{ibata05}
Ibata R., Chapman S., Ferguson A. M.~N., Lewis G., Irwin M., Tanvir N., 2005,
  ApJ, 634, 287

\bibitem[Ibata {et~al.}(2001)]{ibata01}
Ibata R., Irwin M., Lewis G., Ferguson A. M.~N., Tanvir N., 2001, Nature, 412, 49

\bibitem[Ibata {et~al.}(2007)]{ibata07}
Ibata R., Martin N.~F., Irwin M., Chapman S., Ferguson A. M.~N., Lewis G.~F.,
  McConnachie A.~W., 2007, ApJ, 671, 1591

\bibitem[Ibata {et~al.}(1994)]{ibata94}
Ibata R.~A., Gilmore G., Irwin M.~J., 1994, Nature, 370, 194

\bibitem[Ibata {et~al.}(2003)]{ibata03}
Ibata R.~A., Irwin M.~J., Lewis G.~F., Ferguson A. M.~N., Tanvir N., 2003,
  MNRAS, 340, L21

\bibitem[{Ibata {et~al.}(2009)}]{ibata09}
Ibata, R., Mouhcine M., Rejkuba M., 2009, \mnras, 395, 126

\bibitem[Larson (1974)]{larson74}
Larson R.~B., 1974, \mnras, 166, 585

\bibitem[{Malin et al.(1983)}]{malin83}
Malin, D. F., Quinn, P. J., Graham, J. A. 1983, ApJ, 272, L5

\bibitem[Majewski {et~al.}(2002)]{majewski03}
Majewski S. R., et al., 2003, \apj, 599, 1082

\bibitem[Marigo \& Girardi (2007)]{marigo07}
Marigo P., \& Girardi L., 2007, \asa, 469, 239

\bibitem[Mackey \& Gilmore (2004)]{mackey04}
Mackey A.~D., \& Gilmore, G.~F., 2004, \mnras, 355, 504

\bibitem[Mackey et al. (2010)]{mackey10}
Mackey A.~D., et al., 2010, \apj Letter, in press

\bibitem[Marleau et al. (2000)]{marleau00}
Marleau et al. 2000, \aj, 120, 1779

\bibitem[Mouhcine \& Lan\c{c}on (2002)]{ml02}
Mouhcine M., Lan\c{c}on A., 2002, \asa, 393, 149

\bibitem[Mouhcine et al.(2007)]{mouhcine07}
Mouhcine, M., Rejkuba M., Ibata R., 2007, \mnras, 381, 873

\bibitem[Mouhcine et al.(2010a)]{mouhcine10a}
Mouhcine, Ibata R., Rejkuba M., 2010, \apjl, 

\bibitem[Mouhcine et al.(2010b)]{mouhcine10b}
Mouhcine, M., Harris W.~E., Ibata R., Rejkuba M., 2010, \mnras, 

\bibitem[Mould et al. (2000)]{mould00}
Mould, J. R., et al. 2000, \apj, 536, 266

\bibitem[Mould (2005)]{mould05}
Mould J.,  2005, \aj, 129, 698

\bibitem[Newberg {et~al.}(2002)]{newberg02}
Newberg H. J., et al. 2002, \apj, 569, 245

\bibitem[Odenkirchen {et~al.}(2001)]{odenkirchen01}
Odenkirchen M., et al., \apj, 548, 165

\bibitem[Peng {et~al.}(2004)]{peng04}
Peng E.~W., Ford H.~C., Freeman K.~C., 2004, \apj, 602, 685

\bibitem[Rejkuba et al. (2001)]{rejkuba01}
Rejkuba M., Minniti D., Silva D.~R., Bedding T.~R., 2001, A\&A, 379, 781

\bibitem[Rejkuba et al. (2002)]{rejkuba02}
Rejkuba M., Minniti D., Courbin F., Silva D.~R., 2002, \apj, 564, 688

\bibitem[Rejkuba et al. (2003)]{rejkuba03}
Rejkuba M., Minniti D., Silva D.~R., Bedding T.~R., 2003, \asa, 411, 351

\bibitem[Rejkuba et al. (2005)]{rejkuba05}
Rejkuba M., Gregio L., Harris W.~E., Harris G.~L.~H., Peng E.~W., 2005, \apj, 63, 262

\bibitem[Robin et al. (2003)]{robin03} 
Robin A., Reyl\'e C., Derri\`ere S., \& Picaud S., 2003, \asa, 409, 523

\bibitem[Schlegel et al.(1998)]{schlegel98}
Schlegel D.~J., Finkbeiner D.~P., Davis M., 1998, \apj, 525, 500

\bibitem[Soria et al.(1996)]{soria96}
Soria R., et al., 1996, \apj, 465, 79

\bibitem[Stetson (1994)]{stetson94}
Stetson P. B., 1994, \pasp, 106, 250

\bibitem[Stickel et al.(2004)]{stickel04}
Stickel M., van der Hulst J.~M., van Gorkom J.~H., Schiminovich D., Carilli C.~L. 2004,
\asa, 415, 95

\bibitem[Toomre (1977)]{toomre77}
Toomre A., 1977, ARA\&A, 15, 437

\bibitem[VandenBerg et al.(2006)]{vandenberg06}
VandenBerg D.~A., Bergbusch P.~A., Dowler P.~D., Swenson F., 2006, \apjs, 162, 375

\bibitem[Yanny {et~al.}(2000)]{yanny00}
Yanny B., et al., 2000, \apj, 540, 825

\bibitem[Yanny {et~al.}(2003)]{yanny03}
Yanny B., et al., 2003, \apj, 588, 824

\bibitem[Woodley {et~al.}(2010)]{woodley10}
Woodley et al. 2010, ApJ, 708 1335

\bibitem[Zinn (1993)]{zinn93}
Zinn R., 1993, in Smith G.~H., Brodie J.~P., eds. ASP Conf. Ser. 48.
The Globular Cluster-Galaxy Connection. Astron Soc. Pac. San Francisco, p. 38


\bibitem[Zolotov et al. (2009)]{zolotov09}
Zolotov A., Willman B., Brooks A.~M., Governato F., Brook C.~B., Hogg, D.~W., 
Quinn T., Stinson G., \apj, 702, 1058





\end{thebibliography}
\end{document}